\documentclass[10pt,twocolumn]{article}

\usepackage[a4paper,margin=1.8cm]{geometry}
\usepackage{graphicx}
\usepackage{amsmath,amssymb}
\usepackage{bm}
\usepackage{cite}
\usepackage[T1]{fontenc}
\usepackage{lmodern}
\usepackage{authblk}
\usepackage{hyperref}
\title{From quantum emitters to technological platforms: a perspective on semiconductor quantum dots}

\author[1]{Klaus D. Jöns}
\author[2]{Dorian A. Gangloff}
\author[3]{Doris E. Reiter}

\affil[1]{Institute for Photonic Quantum Systems (PhoQS), Center for Optoelectronics and Photonics Paderborn (CeOPP), and Department of Physics, Paderborn University, Warburger Straße 100, 33098 Paderborn, Germany}

\affil[2]{Cavendish Laboratory, JJ Thomson Avenue, Cambridge, CB3 0US, United Kingdom}

\affil[3]{Department of Physics, Condensed Matter Theory, TU Dortmund University, 44221 Dortmund, Germany}

\begin{document}

\maketitle

\abstract{
Photonic quantum technologies are based on controlled photon states. To generate these, quantum dots have emerged as excellent sources of single and entangled photons. In this Perspective, we briefly review the chronological development of quantum dots as photon sources, from their beginnings as quantum emitters in the early 2000s to the advanced photonic devices being developed today. We highlight the interplay between fundamental research and industrial aspects before discussing the photonic quantum technologies we expect to emerge in the near future. }

\section{Introduction}
Non-classical states of light, such as single and entangled photons, form the basis of photonic quantum technologies. With applications ranging from secure quantum communication to photonic quantum computing, they lead to the vision of a quantum internet \cite{kimble2008quantum} and the second quantum revolution. 

The first quantum revolution, rooted in the understanding of quantum mechanics, gave rise to transformative technologies such as the transistor \cite{bardeen1948transistor} and the laser \cite{maiman1960ruby}. These underpin modern electronics, computing, and global communication networks. Solid-state systems, in particular semiconductors, were key to this development, not only because of the underlying physics but also due to their robustness and technological maturity developed over time.

Importantly, this revolution was not driven by fundamental discoveries alone. Its disruptive impact on society came from translating these concepts into precise and reliable devices. This has always been a two-way street, as building and optimising devices exposed gaps in understanding and led to new insights into solid-state systems. For example, semiconductor heterostructures and the quantum cascade laser \cite{faist1994quantum}, as well as the quantum Hall effect \cite{von1986quantized}, revealed new regimes of electron transport and quantum behavior. Over several decades, advances in materials science, fabrication, and system integration, closely linking academic research and industrial development, resulted in increasing levels of control. In this way, abstract quantum principles were not only turned into practical technologies, but also further refined and expanded through their implementation.

As the second quantum revolution unfolds, centered on the controlled use of quantum coherence and entanglement, solid-state platforms, particularly semiconductors with strong light-matter interactions, once again play a central role. In this context, the generation of non-classical states of light becomes a key requirement, with semiconductor quantum dots (QDs) currently representing the most mature and promising platform.

In this Perspective, we briefly review the chronological development of QDs from photon sources to quantum technologies since 2000. We highlight the parallel progress along four key directions: optics, growth, spin, and theory. We discuss the interplay between fundamental and industrial research and conclude by looking into the future, identifying four directions: fundamental questions, applications, scaling, and optimization.

\section{The past}

\begin{figure*}
\centering
    \includegraphics[width=0.9\textwidth]{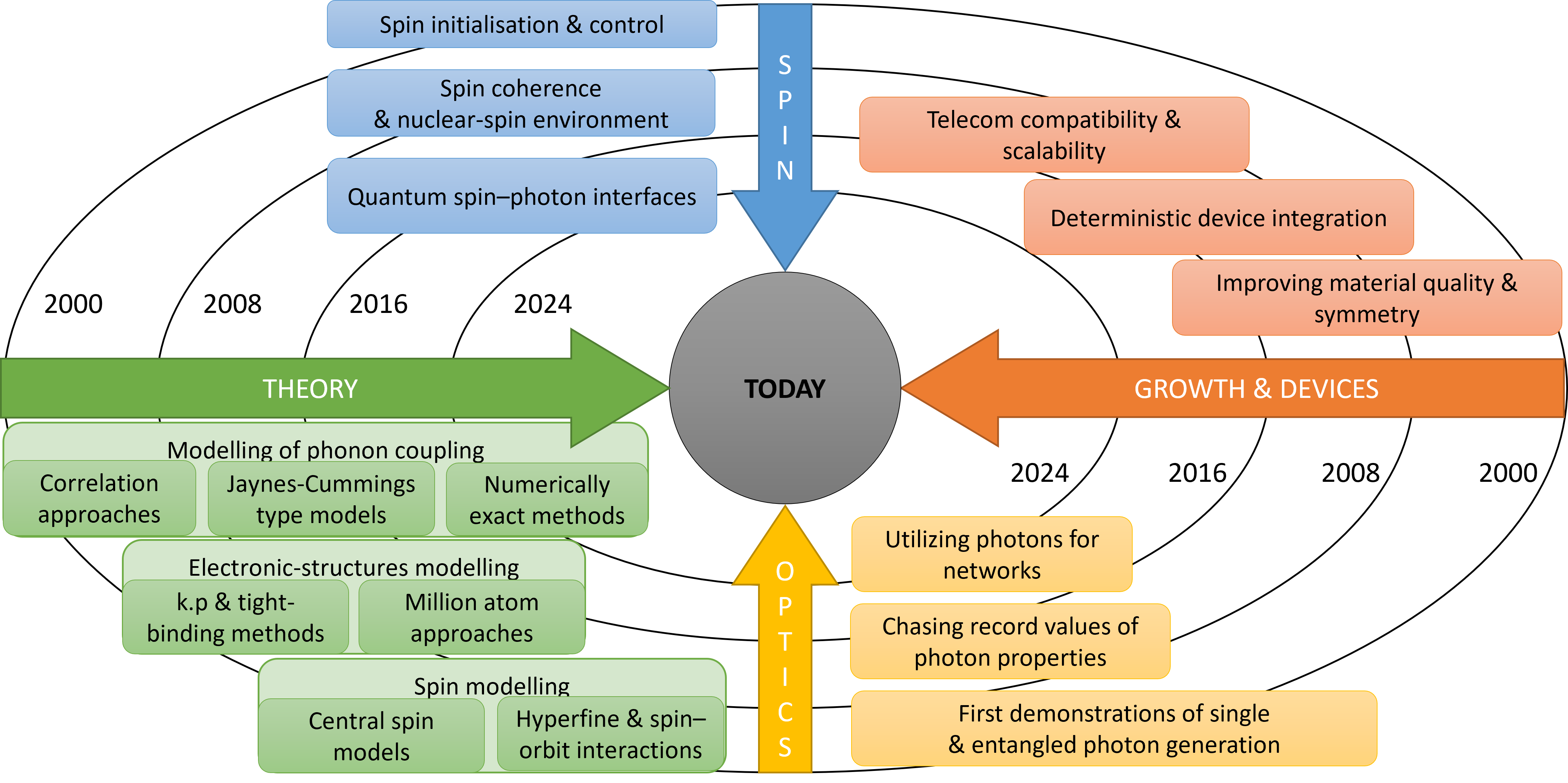}
\caption{\textbf{Illustration of the evolution of the research field of QDs from around 2000 to the present.} The field has progressed along four main axes: optics, growth and devices, theory, and spin.}
\label{fig.past}
\end{figure*}

The concept of semiconductor QDs is more than 50 years old, as recognised by the 2023 Nobel awards \cite{NobelChem2023}. From the 1980s to the early 2000s, the growth of zero-dimensional quantum systems was a focus of fundamental condensed matter physics, establishing the properties of confined semiconductor systems. Building on this research, the field began to explore single, isolated QDs as artificial atoms and their light emission, opening the field of semiconductor quantum photonics around 2000. The developments along the four directions—growth, theory, optics, and spin—are illustrated in Fig.~\ref{fig.past}, highlighting key achievements over the last three decades. The research community addressed closely related questions, with several groups competing toward the same goals. The detailed progress has been covered in many reviews; here, we focus on selected publications to provide a representative account of the field's evolution.

\subsection{\textbf{Optics}}
Major research lines for QDs have been driven by the goal of generating quantum states of light. Using QDs as light-matter sources of non-classical light began in the early 2000s with the first demonstrations of single-photon emission \cite{michler2000quantum, santori2001triggered}. A few years later, QDs were shown to generate entangled photon pairs via the biexciton–exciton cascade \cite{akopian2006entangled}. These developments established QDs as photon sources that, in contrast to spontaneous parametric downconversion (SPDC) sources, provided photons on-demand. In parallel, resonant coherent optical control advanced rapidly, exemplified by the observation of Rabi oscillations \cite{zrenner2002coherent,stufler2006two}.

Throughout the 2010s, the field concentrated on optimizing the key figures of merit for single-photon sources, namely brightness and efficiency, single-photon purity and indistinguishability~\cite{Michler2017QuantumDots}, and the entanglement fidelity of photon pairs \cite{somaschi2016near,huber2017highly}. Excitation schemes were studied to further improve photon generation, for example via chirped pulses \cite{kappe2025chirped} or schemes using auxiliary states, such as phonon-assisted emission~\cite{Luker2019} or multiple pulses \cite{bracht2021swingup}.

Since the early 2020s, the availability of high-quality sources has enabled a shift towards multi-photon experiments and all-photonic networking. QD photons have been distributed across metropolitan-scale links, and teleportation experiments using photons from multiple QDs have been conducted \cite{strobel2025telecom, laneve2025quantum}. A major focus is to build larger-scale photonic entanglement, as useful towards quantum repeaters or quantum computing. This includes the generation of strings of entangled photons or Cluster states \cite{schwartz2016deterministic}, intimately tied to the development of coherent spin qubits in QDs (as discussed below). Another approach towards higher-dimensional states is photon fusion \cite{prasad2025analytical} that relies on perfect single photons to start with. 

\subsection{\textbf{Growth \& Devices}}
The demands for improving key photonic figures of merit was a major driver to advance QD growth.  From the mid-2000s to mid-2010s growth optimization targeted primarily a reduction of the inhomogeneous emission linewidth and the fine-structure splitting (FSS). This involved suppressing charge noise both passively through higher purity growth and actively by integrating QDs into diode structures. The optimizaion of the diode structures took time as they require doping and therefore deliberately introduced impurities. Materials and growth parameters were tuned to enhance dot symmetry, to reduce the FSS, being essential for high-fidelity entanglement generation \cite{Young.Stevenson.ea:2006,Hafenbrak.Ulrich.ea:2007}. Growth concepts explicitly targeting low FSS across different emission wavelengths -- via carefully engineered epitaxy and substrate/material choices -- reinforced the notion that symmetry is a growth parameter, not merely a post-growth tuning problem \cite{SkibaSzymanska2017}.

Then, the research shifted towards determinism, scalability, and control. Pushing towards scalable device manufacturing, the next goal was to place a suitable QD at a desired position relative to a cavity or waveguide mode with high yield, while retaining near-ideal optical properties. Two complementary strategies matured: (i) site-controlled growth with the emitter placed deterministic already at the epitaxy stage~\cite{Joens.Atkinson.ea:2013} and (ii) “locate-and-fabricate” approaches, where high-quality dots are first optically identified, followed by deterministic nanofabrication around them \cite{Dousse.Suffczynski.ea:2010,Sapienza.Davanco.ea:2015}, including active charge control and stabilisation~\cite{Nowak2014}. Another important stream was the active strain tuning of QD emission wavelength, reaching THz scale, and compensation of the FSS~\cite{Rastelli2012, Martin-Sanchez_2018}.
%This was made possible by the coming age of a category of lattice-matched and strain free dots grown by local droplet etching.

In the last decade, telecom compatibility has evolved from a wavelength “nice-to-have” into a hard design constraint for practical, low-loss quantum communication and computing. Growth strategies include bandgap engineering to extend emission into the O- and even C-band, while epitaxy has been developed explicitly for low-FSS and cavity-compatible telecom emission. A representative growth route is droplet epitaxy of InAs/InAlAs dots on vicinal GaAs(111)A, using a largely relaxed InAlAs metamorphic layer while maintaining a flat surface; these dots emit in the O-band with low FSS. On the device side, deterministic fabrication was extended to the O-band by combining in-situ electron-beam lithography with bonding and backside mirrors, together with strain engineering via a strain-reducing layer to redshift InGaAs/GaAs QDs towards O-band emission.

From the late 2010s onwards, droplet-etched/nanohole-based GaAs-family platforms \cite{Heyn.Stemmann.ea:2009} became benchmark emitters because they combine low dot densities, high symmetry, and excellent optical performance, enabling low multi-photon emission, high indistinguishability, and strong entanglement \cite{Huber.Reindl.ea:2017}. Resonant-excitation techniques and improved device environments enabled on-demand, background-free emission~\cite{Schweickert.Joens.ea:2018} and near-unity indistinguishability \cite{Schoell.Hanschke.ea:2019}. These droplet-grown dots also enabled major breakthroughs in spin and spin-photon control.

A unifying trend across these decades is the progressive tightening of constraints: first demonstrating nonclassical emission, then engineering coherence and symmetry, then achieving deterministic integration and telecom compatibility, adding new functionalities by external tuning knobs, and now co-optimizing photons and spins. The next stage will likely be defined by reproducibility at scale—making transform-limited emission, high indistinguishability, low FSS (or robust tuning), and long-lived spin coherence routine across many devices at telecom wavelengths—together with manufacturable photonic integration and packaging.

\subsection{\textbf{Spin}}
In parallel with the other directions, spin physics in epitaxial, optically active QDs formed a distinct research direction in the late 2000s, before recently converging with photonic research into a unified effort on optimized spin–photon interfaces .
 
In the first phase ($\sim$2000–2010), experiments established QDs as coherent light–matter interfaces through the initialization, readout and control of single spins, including high-fidelity spin preparation and coherent spin rotations~\cite{Atature2006,Press2008a}. Together with microsecond-scale hole-spin coherence~\cite{Brunner2009}, these results established the basis for spin-photon interfaces. Simultaneously, single-electron spins were shown to interact strongly with the nuclear-spin ensemble, resulting in electron-spin decoherence but also revealing the nuclear spins as a controllable for improving coherence through optical control of the hyperfine interaction~\cite{Lai2006,Greilich2006}.

In the second epoch, the field progressed towards spin-photon and heralded distant spin-spin entanglement~\cite{Gao2012,DeGreve2012,Stockill2017a}, while revealing limited spin coherence as a key bottleneck. This motivated extensive studies of the hyperfine interaction, with ultrafast spin control enabling the electron spin to probe its native nuclear-spin environment and revealing strain-induced nuclear inhomogeneities as a fundamental limit to electron-spin coherence~\cite{Bechtold2015}. Nuclear-spin narrowing and spin control further revealed hyperfine feedback mechanisms and established the nuclear environment as a system that could be actively manipulated to improve coherence~\cite{Ethier-Majcher2017, Gangloff2019}. Comparisons with alternative platforms such as NV centers highlighted the trade-off between the superior optical interfaces of QDs and their more limited spin coherence~\cite{Atature2018}.

The most recent epoch reflects a shift from identifying limitations to actively overcoming them. Nuclear-spin cooling and phase-coherent spin control enabled high-fidelity manipulation of electronic spins and active control of the nuclear-spin environment~\cite{Gangloff2019,Bodey2019,Jackson2022}. In parallel, the strain-induced limits to spin coherence were overcome by materials design: strain-free GaAs QDs enabled substantially extended electron-spin coherence through dynamical decoupling~\cite{Zaporski2023_natnano}. In these low-inhomogeneity nuclear-spin systems, nuclear-spin cooling by quantum-algorithmic feedback~\cite{Jackson2022} further increased the electron-spin coherence time $T_{2}^{*}$ by over two orders of magnitude to $0.6\mu$s~\cite{Nguyen2023}, while direct NMR control of the nuclear-spin ensemble enabled high-fidelity dynamical decoupling to 100\,ms~\cite{Dyte2025}. The symmetry of GaAs QDs has also enabled tailoring of their optical and magnetic properties prior to growth, an important step towards scalable devices~\cite{Schimpf2025optical}. This broad progress has enabled the long-sought transfer of quantum information from an electron spin qubit to a nuclear spin-wave (magnon) mode~\cite{Appel2025}.

\subsection{\textbf{Theory}}
Aligned to the advances in QD optics, growth, and spin, the theory has shown great progress.

Light–matter interaction has long been studied in atomic systems, leading to Jaynes-Cummings-type models \cite{larson2021jaynes}. In contrast, as solid-state nanostructures, QDs are inherently coupled to their semiconductor environment, most prominently through electron–phonon interactions. With their discrete energy structure, the question of a phonon bottleneck was already discussed from the 1990s onwards \cite{bockelmann1990phonon,li1999phonon}. In the limit of treating the QD as a two-level system, diagonal phonon coupling, often referred to as a pure-dephasing Hamiltonian~\cite{mahan2000many}, gives rise to non-Markovian dynamics, visible in phonon-sideband emission~\cite{krummheuer2002theory}.

In the early 2000s, theory tackled phonon-induced decoherence during coherent excitation of QDs, using techniques such as correlation expansion~\cite{krugel2005role}. A notable prediction in 2007 was the reappearance of Rabi oscillations at strong driving fields~\cite{vagov2007nonmonotonic}, only recently confirmed experimentally~\cite{hanschke2025experimental}. In the 2010s, theoretical treatments extended the Jaynes–Cummings framework to include phonons~\cite{kaer2010non}, with polaron master equations and numerically exact path-integral methods \cite{roy2012polaron,vagov2011real}. From the 2020s onwards, the field has increasingly moved towards tensor-network approaches combining numerical exactness with improved computational efficiency~\cite{fux2024oqupy,cygorek2022simulation}. Today, QD dynamics including phonon coupling, radiative decay and losses can be simulated numerically exactly, including multi-time correlation functions and state tomography \cite{bracht2026tunable}.

In parallel, the modelling of the electronic structure of QDs advanced through methods including million-atom approaches~\cite{zielinski2010atomistic} and tight-binding methods~\cite{gawarecki2019importance}, accounting for Coulomb correlations, strain effects, and alloy decomposition. In particular, higher excited states that might become relevant for advanced optical protocols are highly complex and call for advanced theoretical descriptions~\cite{henzler2021femtosecond}, as does the understanding of spin coupling and $g$-factors~\cite{Schimpf2025optical}.

Likewise, theoretical descriptions of hyperfine and spin-orbit interactions in QDs have advanced greatly. The interaction of an electron spin with the nuclear-spin ensemble was described using the well-known central-spin model, applied to spin relaxation and decoherence~\cite{Merkulov2002,Khaetskii2002}, leading to proposals for storing and manipulating quantum information in the nuclear-spin ensemble~\cite{Taylor2003a}. Increasingly accurate many-body calculations described spin coherence~\cite{Yao2006,Cywinski2009a}, while semiclassical approaches captured feedback dynamics of the Overhauser field and nuclear-spin narrowing~\cite{Yang2013,Barnes2011a}. Incorporating strain inhomogeneities further clarified the role of nuclear-spin disorder and enabled theoretical descriptions of active nuclear-spin control~\cite{Denning2019}. This work ultimately led to proposals and experimental realisations of programmatic nuclear-spin cooling~\cite{Jackson2022} and quantum algorithms for preparing non-Gaussian and error-correcting states of nuclear spins~\cite{Zaporski2023_prx,Franke2026ASpins}.

\section{Interplay Industry and Research} 
\begin{figure}
    \includegraphics[width=\columnwidth]{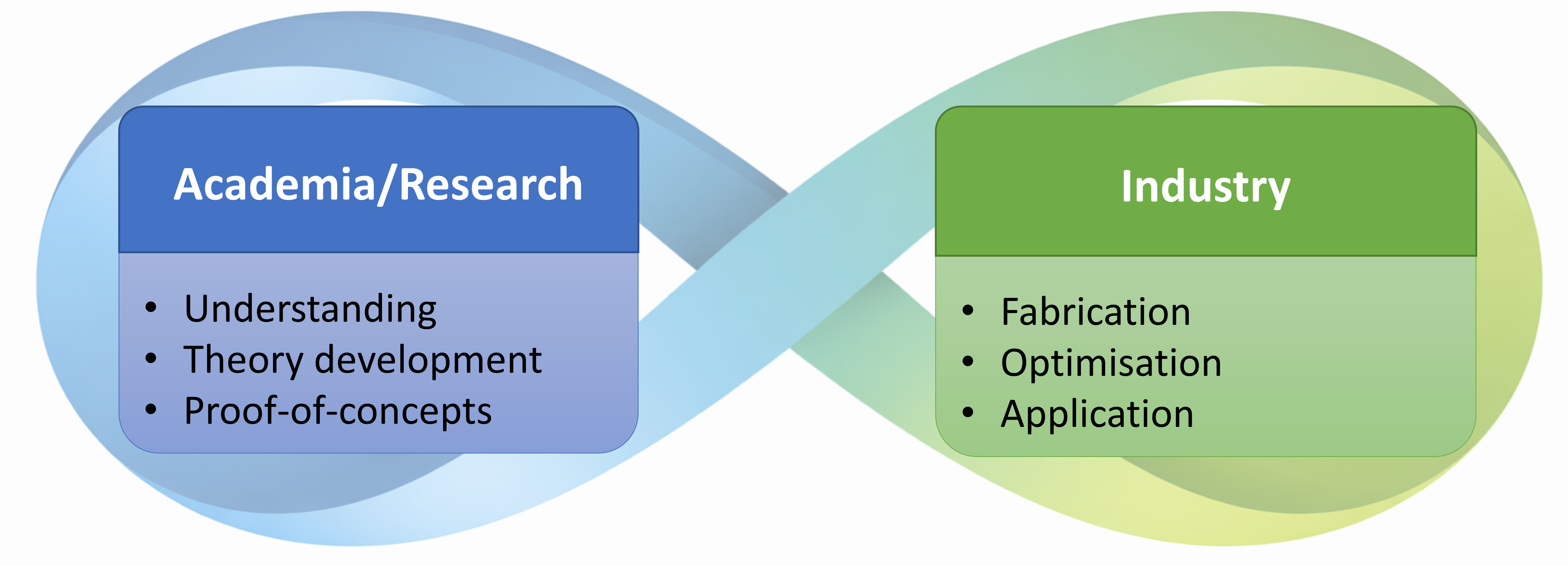}
\caption{\textbf{Visualization of the feedback loop between academia and industry.} Fundamental experimental and theoretical research enables the development of quantum photonic devices, which are then translated into industrial technologies. Industrial constraints and requirements pose new questions to academia, forming a cycle of innovation.}
%\label{fig.1}
\end{figure}

Industry engagement in deterministic quantum light sources based on QDs is accelerating rapidly, reflecting a broader transition of the field from fundamental research toward early-stage commercialization. A growing number of companies, such as Quandela, Sparrow Quantum, AegiQ, and Saqura, are investing in this technology, motivated by its unique capability to deliver on-demand, high-quality single photons that are widely recognised as essential for scaling photonic quantum technologies. Quantum computing is currently a significant near-term market, given its demanding performance requirements and willingness to adopt enabling hardware at an early stage. In this context, deterministic QD sources offer a clear pathway toward reducing resource overheads and improving scalability in multi-photon architectures.

In our view, this convergence of scientific maturity and industrial interest marks a particularly exciting phase for the field. At the same time, the increasing commercial traction highlights the continued importance of academic research. Many of the remaining challenges—ranging from materials optimization and spin coherence to yield, reproducibility, wafer-scale fabrication, and system integration—are fundamentally rooted in questions that require sustained scientific investigation. Moreover, academia plays a central role in training the highly specialised workforce that this emerging ecosystem depends on, spanning expertise in quantum optics, semiconductor growth, nanofabrication, and device engineering.

Looking ahead, the most promising developments will likely emerge from a close interplay between academia and industry. While companies are driving system-level integration and application-oriented progress, academic research remains essential for addressing foundational bottlenecks and enabling the next generation of breakthroughs. Taken together, these dynamics suggest that the field is entering a phase of unprecedented opportunity, where advances in science, engineering, and commercialization reinforce each other—positioning QD single-photon sources as a cornerstone technology for future quantum infrastructures.

\section{The future}

\begin{figure*}
\centering
    \includegraphics[width=0.9\textwidth]{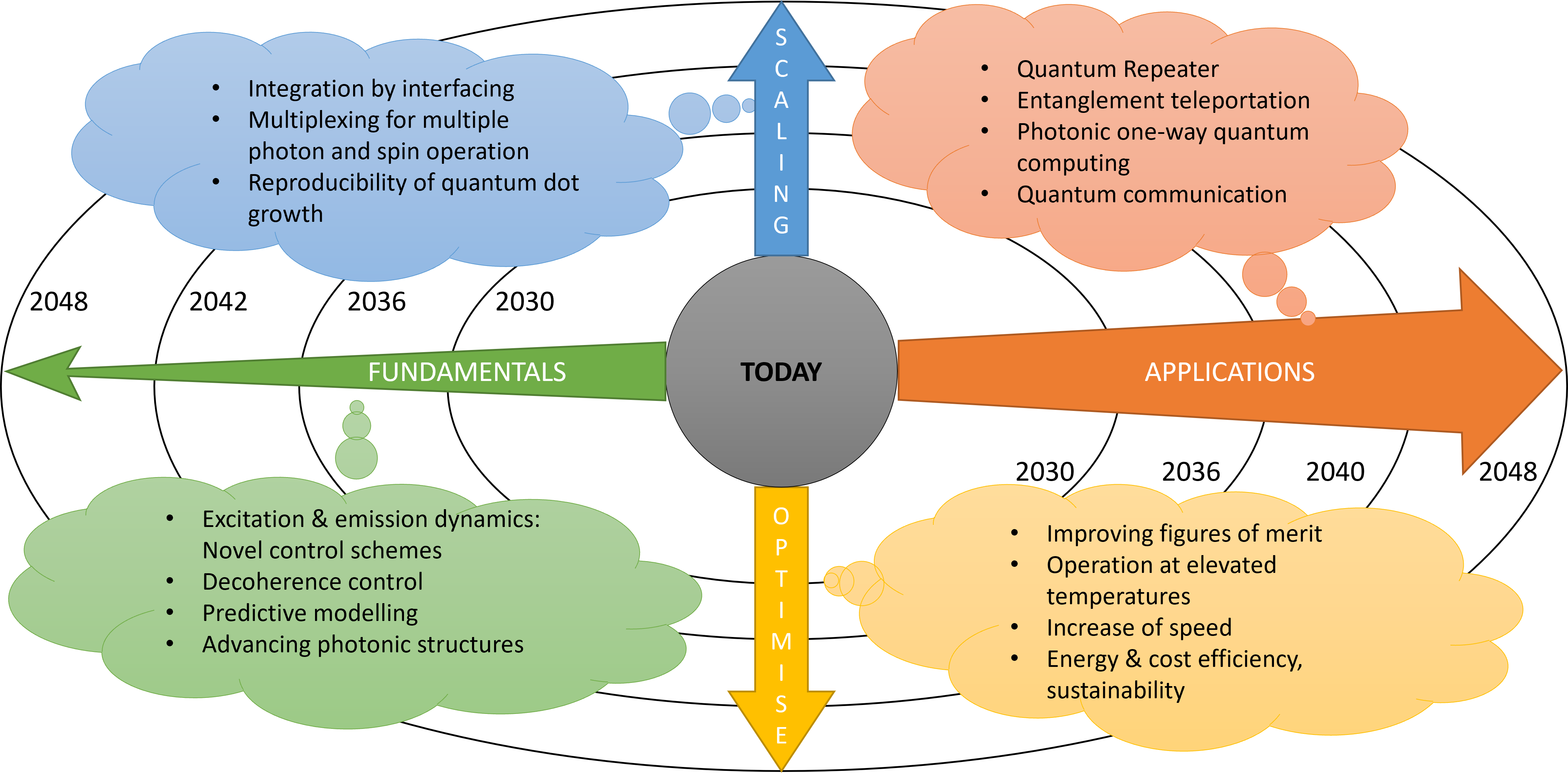}
\caption{\textbf{Illustration of the evolution of the research field of QDs towards the future.} We envision along four lines: fundamentals, applications, scaling and optimization.}
\label{fig:future}
\end{figure*}

How the field may evolve from the present state (2026) is illustrated in Fig.~\ref{fig:future}. We organise future developments along four main directions: fundamentals, applications, scaling, and performance. While efforts in scaling and performance will continue to grow as the field matures, we expect a gradual shift from fundamental studies toward the development of applications, as indicated by the size of the arrows.
\subsection{\textbf{Fundamentals}} 
Although the main concepts of solid-state quantum emitters are well established and many proof-of-concept demonstrations have been achieved, numerous open questions remain from both theoretical and experimental perspectives. New effects continue to emerge, such as novel excitation schemes such as the SUPER approach \cite{bracht2021swingup,karli2022super}, or the effect of excitation processes on entanglement fidelity \cite{seidelmann2022two,basso2023signatures}. At the same time, due to the increasing maturity of the field, further progress is expected to come less from the discovery of fundamental effects and more from gaining a deeper understanding of the interplay of effects and their control.

A comprehensive theoretical description remains challenging. Bridging predictive atomistic modelling with the dynamics of excitation and emission and the influence of the solid-state environment remains an open problem and is essential for a complete understanding of these systems. This should ultimately include the detection process, as signals in multi-time correlation functions can depend strongly on detection time and other parameters. The development of interfaces and coupling to photonic structures goes beyond purely engineering challenges. New concepts for efficiently routing photons, controlling light-matter interaction, and minimizing losses will be crucial for the realization of practical devices. Here, significant conceptual advances are still required.

\subsection{\textbf{Optimizations}} 
Optimization and a high level of control are prerequisites for any application, and there is still significant room for improvement. The most direct progress will come from improving the key figures of merit of QD photon sources. For applications such as photonic quantum computing, minimum thresholds for photon quality must be reached, depending on the specific protocol. Achieving these requirements is closely linked to a deeper understanding and control of decoherence mechanisms. This understanding will also help to reach elevated operation temperatures for QDs \cite{lehner2023beyond,bracht2023temperature}.

At the same time, efforts to mitigate decoherence can also lead to new opportunities. A prominent example is the role of phonons, which are typically considered a major source of decoherence, but can be exploited in phonon-assisted excitation schemes to enable off-resonant excitation with improved performance \cite{cosacchi2019emission,Reindl2019}.

Beyond these core metrics, additional system-level properties become increasingly important for the integration of photon sources into practical devices. One key parameter is the operational speed, determined by the repetition rate of photon emission. Increasing repetition rates requires both faster emission dynamics and advances in detection technologies. While, in principle, these aspects are scalable, the interplay between excitation and emission at high rates may give rise to new physical effects that need to be understood.

Another important consideration is sustainability, including aspects such as energy efficiency, fabrication cost, and long-term stability. These factors will play an increasingly important role as quantum technologies move from laboratory demonstrations toward real-world deployment.

\subsection{\textbf{Scaling}}
Scaling up QD-based systems can be broken down into three sub-challenges: integration, multiplexing, and reproducibility. Addressing each in turn, while minimizing photon loss below the required thresholds~\cite{Wein2025,Chan2026}, is key to enabling the multi-photon protocols for photonic quantum computing and networks, whether fusion-based or hybrid spin-photon architectures.

The main integration challenges is to interface QD sources grown in III-V materials with the electro-optics hosted in materials with strong nonlinearities. This is required for path, phase, and polarisation control of individual photons. Photonic integrated circuits (PICs), in turn, provide low-loss, CMOS-compatible platforms for interferometers and detectors required for two-photon operations. These components are typically realised in different material platforms, making heterogeneous integration an important challenge. A promising approach is the modularisation of QD emitters coupled to optical fibres, for example through waveguide structures connected to tapered fibres~\cite{Sipahigil2016} or fibre-pigtailed micropillar structures~\cite{Margaria2025}. Reducing insertion losses through optimised mode-matching remains an important engineering challenge. The operational wavelength of the sources imposes strict constraints on device dimensions and absorption losses, and may thus steer QD sources at different ends of the wavelength spectrum towards different integration strategies.

The second major challenge towards scaling up is multiplexing on multiple scales. On the single QD level, controlling multiple spins (e.g. molecular or nuclear) can enable the generation of multi-dimensional photonic entanglement at the source~\cite{Buterakos2017} or the buffering of entanglement generation in a network by a memory~\cite{Appel2025}. At the scale of multiple dots, collective excitation \cite{kappe2023collective} or path multiplexers~\cite{Valencia2026} will enable a key strength of spin-optical systems, namely programmable connectivity (ultimately all-to-all), which can significantly reduce error-correction overheads~\cite{Wein2025}. At the level of a quantum processing unit (QPU), packing density within available cryogenic footprints will become important, thereby demanding multiplexing of both quantum and classical channels to and from the QPU.

Reproducibility is the last key to scaling: the optical and spin properties of QDs need to be predictable within reduced tolerances prior to growth, with sufficiently high yield to produce the hundreds to thousands of identical sources needed for fault-tolerant systems. This will require a combination of growth standardisation and high-throughput characterisation. Dedicated epitaxy reactors tailored to specific QD structures can improve control and reproducibility of growth conditions. Downstream, automated wafer-wide characterisation of wavelength, brightness, linewidth, and spin lifetime will enable the identification and processing of QDs that meet the required specifications.

\subsection{\textbf{Applications}}
Over the next decade, we expect QDs to transition from high-performance laboratory systems to integral components of deployed quantum technologies. In the near term, their role will likely consolidate in applications that demand the highest source quality, such as photonic quantum computing based on cluster states and advanced quantum optics experiments. As coherence properties, device reproducibility, and nanophotonic integration improve, QDs could enable quantum repeater schemes and act as coherent quantum nodes in networked architectures. Their deployment as reliable sources for entanglement distribution and quantum key distribution in metropolitan fiber networks also appears increasingly realistic as telecom-compatible platforms mature. Another possible application is quantum transduction, transferring states, for example, from the microwave regime to telecommunication frequencies.

A key milestone along this trajectory will be the shift from standalone components to fully integrated, application-specific systems. Within this framework, QDs will serve as an embedded hardware layer inside photonic quantum processors or turnkey quantum communication links, rather than as individual sources operated by specialists. Achieving this will require continued progress in scalability, packaging, system integration, standardisation, and long-term operational stability. If these challenges are addressed, QDs could become a central technology in quantum computing and quantum communication infrastructures within the coming decade.

\section{Summary}
All this leads to the question: are solid-state quantum emitters still a worthwhile field to enter? The answer is clearly yes, but one should set their expectations right: As the field matures, you may find yourself in a few years working in an industrial setting rather than on the theoretical or experimental foundations of physics. At the same time, there is still much to explore and to learn about the fascinating aspects of quantum mechanics in these systems. It is also possible that entirely new research directions will emerge from solid-state quantum emitters in the future. For these reasons, we strongly encourage readers to join this spirited community.

\section{Acknowledgments}
D.E.R. acknowledges funding from the DFG under projects BRAIDS (560765700) and Photon-MagiQ (577681036), and from the BMFTR within the QuantERA project MEEDGARD (16KIS2058). K.D.J. acknowledges funding from the BMFTR through projects QR.N (16KIS2206) and TuF-ToPiQC (13N17238) and from the ERC under the EU's Horizon 2020 research and innovation programme (LiNQs 101042672). D.A.G. acknowledges support from a Royal Society URF.

\bibliographystyle{unsrt}
\bibliography{EPL}

\end{document}